\documentclass[12pt,a4paper]{article}
\usepackage[utf8]{inputenc}
\newcommand\tab[1][1cm]{\hspace*{#1}}
\usepackage{amsmath}
\usepackage{amsfonts}
\usepackage{amssymb}
\usepackage{graphicx}
\usepackage{cite}
\usepackage{authblk}
\usepackage{hyperref}
\usepackage{float}
\begin{document}
\author{Ph.Wilina\footnote{wilina.phd.phy@manipuruniv.ac.in} $^{1}$,
M.Shubhakanta Singh\footnote{mshubhakanta@yahoo.com} $^{1}$ and N.Nimai Singh\footnote{nimai03@yahoo.com} $^{1,2}$\\
$^{1}$\small Department of Physics, Manipur University, Imphal-795003, India\\ 
$^{2}$\small Research Institute of Science and Technology,  Imphal-795003, India}
\date{}
\title{Deviations from Tribimaximal and Golden Ratio mixings under radiative corrections of neutrino masses and mixings.}
\maketitle
\begin{abstract}
The impact of renormalization group equations(RGEs) on neutrino masses and mixings at high energy scales in Minimal Supersymmetric Standard Model(MSSM) is studied using two different mixing patterns such as  Tri-Bimaximal(TBM) mixing and Golden Ratio(GR) mixing in consistent with cosmological bound of the sum of three neutrino masses, $\sum _{i}|m_{i}|$. Magnifications of neutrino masses and mixing angles at low energy scale, are obtained by giving proper input masses, and mixing angles from TBM mixing matrix and GR mixing matrix at high energy scales. High energy scales, $M_{R}$ such as $10^{13}$GeV,$10^{14}$GeV,$10^{15}GeV$ are employed in the analysis. The large solar($\theta_{12}$) and atmospheric($\theta_{23}$) neutrino mixing angles with zero reactor angle ($\theta_{13}$) from both TBM mixing matrix and GR mixing matrix at high scale, can magnify the reactor angle($\theta_{13}$) at low energy scale in 3$\sigma$ confidence level. Both cases of normal hierarchy(NH) and inverted hierarchy(IH
) are addressed here. In normal hierarchical case, it is found that $\theta_{23}\simeq51.1^{\circ}$ and that in inverted hierarchical case is $\theta_{23}\simeq39.1^{\circ}$ in both mixing patterns.  Possibility of  $\theta_{23}>45^{\circ}$ or $\theta_{23}<45^{\circ}$ is observed at low scale. The analysis shows the validity of the two mixing patterns at high energy scale.$\newline$
\tab \textbf{Keywords}: renormalization group equations, Minimal Supersymmetric Standard Model(MSSM), Tribimaximal(TBM) mixing, Gol-den ratio(GR)mixing.
\end{abstract}

\section{Introduction}
The renormalization group equations(RGEs)\cite{Das_2005,Xing_2018,Chao_2007} in neutrino physics serve as an essential way to magnify neutrino mixing angles($\theta_{12},\theta_{23},\theta_{13}$) from high energy scales to low energy scale. In order to obtain a radiative correction to the neutrino masses  and the three mixing angles, the RGEs are made to run in terms of neutrino mass eigenvalues and sine of the mixing angles at different high energy scales, $M_{R}$($10^{13}GeV,10^{14}GeV,10^{15}GeV$). The TBM mixing\cite{Xing_2002,Harrison_2002,Ahn_2012} which was regarded as a faithful candidate of the PMNS mixing\cite{dattoli2008neutrino}, is ruled out after the discovery of $\theta_{13}\neq{0}$\cite{Ahnn,King:2011xu}. The observation of $\theta_{13}\neq{0}$ has made an important contribution in $\delta_{CP}$ in leptonic sector. Both the TBM mixing and GR mixing\cite{Feruglio_2011,Kajiyama_2007,Ding_2012} predict exact $\theta_{13}=0$. However, there is a possibility of obtaining non-zero $\theta_{13}$ at low scale by giving mixing angles from TBM mixing and GR mixing as input with proper choices of $ m_{i}$(i=1,2,3), consistent with cosmological bound on  $\sum_{i} m_{i}$.\newline

\tab In the work, we are dealing mostly on the validity of the TBM and GR mixing matrix in both normal hierarchy(NH) and inverted hierarchy(IH) at high energy scale along with radiative magnification of $\theta_{13}$ at low energy scale. To obtain these, we consider the evolution equations of neutrino masses and sine of mixings taking into account the scale dependent vacuum expectation value(vev). We consider large $tan \beta$ value which in fact gives a large contribution in obtaining precise values of mixing angles particularly in the magnification of $\theta_{13}$. We notice that with small $tan \beta$ value, radiative magnification of $\theta_{13}$ is not possible. We assign neutrino mass eigenvalues in the form ($m_{1},-m_{2},m_{3}$) with magnitude of $m_{2}$$>$$m_{1}$ and $m_{3}\neq 0$. With very small $m_{3}$ value in IH, the value of $\theta_{13}$ cannot be magnified at low scale.  An updated global analysis of neutrino oscillation measurements provided by NuFit is given in \cite{Esteban_2020}.\newline

\tab  The paper is organised as follows. In section 2, we briefly outline the lepton mixing including the Tribimaximal mixing and Golden Ratio mixing. In section 3, we present the renormalization group equations for neutrino mass eigenvalues and mixing angles. The numerical analysis and results are presented in section 4. We summarise our results in section 5. \newline


\newpage
\section{Tribimaximal mixing and Golden ratio mixing}
 The PMNS mixing matrix
 \begin{equation}
 U_{PMNS}=
 \begin{pmatrix}
 U_{e1} & U_{e2} & U_{e3} \\
 U_{\mu 1} & U_{\mu 2} & U_{\mu 3} \\
 U_{\tau 1} & U_{\tau 2} & U_{\tau 3}\\ 
 \end{pmatrix}
\end{equation} 
 after parametrisation in terms of three rotations R($\theta_{12}$),R($\theta_{13}$),R($\theta_{23}$), is given by\cite{Ii,Pal_2017,Giganti_2018}
 \begin{equation}
 U_{PMNS}=
 \begin{pmatrix}
 c_{13}c_{12} & c_{13}s_{12} & s_{13} \\
 -c_{23}s_{12}-c_{12}s_{13} & c_{12}c_{23}-s_{12}s_{13}s_{23} & c_{13}s_{23}\\
 s_{12}s_{23}-c_{12}s_{13}c_{23} & -c_{12}s_{23}-c_{23}s_{13}s_{12} & c_{13}c_{23}
 \end{pmatrix}
 ,
 \end{equation}
 where $s_{ij}=sin \theta_{ij}$, $c_{ij}=cos \theta_{ij}$. Here, CP violating Dirac $\delta_{CP}$ phase is neglected for simplicity.\newline
 
 The Tri-bimaximal mixing matrix is given by\cite{MA_2010}
 \begin{equation}
 U_{TBM}=
 \begin{pmatrix}
 \sqrt{\dfrac{2}{3}} & \dfrac{1}{\sqrt{3}} & 0 \\
 -\dfrac{1}{\sqrt{6}} & \dfrac{1}{\sqrt{3}}& \dfrac{1}{\sqrt{2}} \\
 \dfrac{1}{\sqrt{6}} & -\dfrac{1}{\sqrt{3}} & \dfrac{1}{\sqrt{2}}
 \end{pmatrix}
 ,
 \end{equation} 
which predicts that $s_{13}^{2}$=0, $s_{23}^{2}$=1/2 and $s_{12}^{2}$=1/3.\newline

 Another approachable pattern of neutrino mixing matrix deals with the golden ratio $\phi$=$\dfrac{1+\sqrt{5}}{2}$, with the assumption tan $  \theta_{12}=1/\phi$, $s_{13}^{2}=0$ and $s_{23}^{2}=1/2$. The golden ratio mixing matrix is thus given by\cite{King}  \newline
 \begin{equation*}
 U_{GR}=
\begin{pmatrix}
 \dfrac{\phi}{\sqrt{2+\phi}} & \dfrac{1}{\sqrt{2+\phi}} & 0 \\
 -\dfrac{1}{\sqrt{4+2\phi}} & \dfrac{\phi}{\sqrt{4+2\phi}} & \dfrac{1}{\sqrt{2}} \\
 \dfrac{1}{\sqrt{4+2\phi}} & -\dfrac{\phi}{\sqrt{4+2\phi}} & \dfrac{1}{\sqrt{2}}
\end{pmatrix}
\end{equation*}
 
\begin{equation}
 \tab      =
 \begin{pmatrix}
 0.850639 & 0.525735 & 0 \\
 -0.371750 & 0.601491 & 0.707107 \\
 0.371750 & -0.601491 & 0.707107
 \end{pmatrix}
 .
 \end{equation}
In both TBM and GR mixing matrices, it is predicted that $\theta_{13}$=0. However, recent neutrino oscillation data denied $\theta_{13}$ to be  zero.\newline
 
In the present paper, we analyse how the TBM mixing and GR mixing at high energy scale of $10^{13}GeV,10^{14}GeV$,  $10^{15}GeV$, can magnify the $\theta_{13}$ value to a required value at low energy scale  along with other necessary neutrino parameters such as $\theta_{12},\theta_{23}$,$\Delta m_{12}^{2}$,$\Delta m_{23}^{2}$. While doing this, the present work also focuses on the consistency with the cosmological bound on $\sum_{i} |m_{i}|$ where $i=1,2,3$.
 \section{RGEs for neutrino mass eigenvalues and mixing angles}
 The neutrino mass matrix taking into account the running of vev is given by\cite{singh2001effects} 
 \begin{equation*}
 m_{i}(t)=v_{u}^{2}(t)K_{i}(t)
 \end{equation*}
 or 
 \begin{equation}
  m_{i}(t)   = v^{2}\dfrac{tan^{2} \beta}{(1+tan^{2} \beta )}K_{i}(t) ,
 \end{equation}
where t=ln($\mu$/1GeV), $v_{u}=vsin \beta$ with $v$=174GeV, $tan \beta$ is the ratio of the two vev's of two Higgs doublets in MSSM and $K_{i}$ is coefficient of dim-5 neutrino mass operator in scale-dependent manner.\newline
 This implies that
\begin{equation}
\frac{d(ln m_{i})}{dt}=\frac{d(ln K_{i})}{dt}+2\frac{d(ln v_{u})}{dt} ,
\end{equation}
where the evolution equations for $K_{i}$ and $v_{u}$ in MSSM are given by
\begin{equation}
\dfrac{dK_{i}}{dt}=\dfrac{1}{16\pi^{2}}\sum_{f=e,\mu,\tau}\Bigg[\Bigg(-\dfrac{6}{5}g_{1}^{2}-6g_{2}^{2}+6Tr(h_{u}^{2})\Bigg)+2h_{f}^{2}U_{fi}^{2}\Bigg]K_{i} ,
\end{equation}
\begin{equation}
\frac{d(v_{u})}{dt}=\dfrac{1}{16\pi^{2}}\Bigg[\dfrac{3}{20}g_{1}^{2}+\dfrac{3}{4}g_{2}^{2}-3h_{t}^{2}\Bigg]v_{u} ,
\end{equation}
where $g_{1},g_{2}$ are gauge couplings, and $h_{u}$, $h_{t}$ are up-quark and top-quark Yukawa coupling respectively.\newline
Now, the RGEs for neutrino mass eigenvalues are given by
\begin{equation}\label{e1}
\frac{d}{dt}m_{i}=\dfrac{1}{16\pi^{2}}\Bigg[-\dfrac{9}{10}g_{1}^{2}-\dfrac{9}{2}g_{2}^{2}+2h_{\tau}^{2}U_{\tau i}^{2}\Bigg]m_{i} ,
\end{equation}
where $h_{\tau}$ is tau-lepton Yukawa coupling.\newline
Also, the evolution equations for the PMNS matrix elements $U_{fi}$ are given by
\begin{equation}\label{e2}
\frac{d}{dt}U_{fi}=-\dfrac{1}{16\pi^{2}}\sum_{k\neq i}\dfrac{m_{k}+m_{i}}{m_{k}-m_{i}}U_{fk}(U^{T}H_{e}^{2}U)_{ki} ,
\end{equation}
where $f=e,\mu,\tau$ ; $i,k$=1,2,3 respectively and $H_{e}$ is the Yukawa coupling matrices of the charged-leptons in the diagonal basis.\newline
Now, we have \newline

$(U^{T}H_{e}^{2}U)_{12}$=$h_{e}^{2}(U_{1e}^{T}U_{e2})$ + $h_{\mu}^{2}(U_{1\mu}^{T}U_{\mu 2})$+ $h_{\tau}^{2}(U_{1\tau}^{T}U_{\tau 2})$ \newline

$(U^{T}H_{e}^{2}U)_{13}$=$h_{e}^{2}(U_{1e}^{T}U_{e3})$ + $h_{\mu}^{2}(U_{1\mu}^{T}U_{\mu 3})$+ $h_{\tau}^{2}(U_{1\tau}^{T}U_{\tau 3})$
\newline

$(U^{T}H_{e}^{2}U)_{23}$=$h_{e}^{2}(U_{2e}^{T}U_{e3})$ + $h_{\mu}^{2}(U_{2\mu}^{T}U_{\mu 3})$+ $h_{\tau}^{2}(U_{2\tau}^{T}U_{\tau 3})$
\newline
As $h_{\tau}>>h_{e},h_{\mu}$, we neglect $h_{e},h_{\mu}$. Then, the above equations become \newline

$(U^{T}H_{e}^{2}U)_{12}\approx h_{\tau}^{2}(U_{1\tau}^{T}U_{\tau 2})$,
$(U^{T}H_{e}^{2}U)_{13}\approx h_{\tau}^{2}(U_{1\tau}^{T}U_{\tau 3})$,\newline

$(U^{T}H_{e}^{2}U)_{23}\approx h_{\tau}^{2}(U_{2\tau}^{T}U_{\tau 3})$
\newline
Denoting $\dfrac{m_{k}+m_{i}}{m_{k}-m_{i}}=A_{ki}$ and using the above three equations together with eqns.\ref{e2}, we have\newline
\begin{equation}
\frac{dU_{e2}}{dt}\approx -\dfrac{1}{16\pi^{2}}\big[U_{\tau 2}h_{\tau}^{2}(A_{32}U_{e3}U_{3\tau}^{\dagger}+A_{12}U_{e1}U_{1\tau}^{\dagger}) ,
\end{equation}
\begin{equation}
\frac{dU_{e3}}{dt}\approx -\dfrac{1}{16\pi^{2}}\big[U_{\tau 3}h_{\tau}^{2}(A_{13}U_{e1}U_{1\tau}^{\dagger}+A_{23}U_{e2}U_{2\tau}^{\dagger}) ,
\end{equation}
\begin{equation}
\frac{dU_{\mu 3}}{dt}\approx -\dfrac{1}{16\pi^{2}}\big[U_{\tau 3}h_{\tau}^{2}(A_{13}U_{\mu 1}U_{1\tau}^{\dagger}+A_{23}U_{\mu }U_{2\tau}^{\dagger}) .
\end{equation}
Upon solving the above three equations using eqn.(2), we have \newline
\begin{equation}\label{e3}
\begin{aligned}
\frac{ds_{12}}{dt}\approx \dfrac{1}{16\pi^{2}}h_{\tau}^{2}c_{12}\big[c_{23}s_{13}s_{12}U_{\tau 1}A_{31}-c_{23}s_{13}c_{13}U_{\tau 2}A_{32}+ \\
U_{\tau 1}U_{\tau 2}A_{21}\big],
\end{aligned}
\end{equation}
\begin{equation}\label{e4}
\frac{ds_{13}}{dt}\approx \dfrac{1}{16\pi^{2}}h_{\tau}^{2}c_{23}c_{13}^{2}\big[c_{12}U_{\tau 1}A_{31}+s_{12}U_{\tau 2}A_{32}\big] ,
\end{equation}
\begin{equation}\label{e5}
\frac{ds_{23}}{dt}\approx \dfrac{1}{16\pi^{2}}h_{\tau}^{2}c_{23}^{2}\big[-s_{12}U_{\tau 1}A_{31}+c_{12}U_{\tau 2}A_{32}\big].
\end{equation}
The equations \ref{e1},  \ref{e3},  \ref{e4} ,  \ref{e5}  are used for numerical analysis in this paper.

\section{Numerical Analysis and Result}
For numerical analysis, different high scales such as $10^{13}GeV,10^{14}GeV,10^{15}GeV$ and  $tan \beta$=58 are considered. Corresponding to these different scales and $tan \beta$, different values of gauge couplings and the third-family Yukawa couplings are taken. Running of these couplings in MSSM is done by bottom-up method i.e from top-quark mass scale, $m_{t}$ (low energy scale) to $M_{R}$(high energy scale) where the B-L symmetry breaks down. For simplicity, we consider $t_{o}=\ln  m_{t}$ where $m_{t}$ is mass of top-quark and $t_{R}=\ln  M_{R}$. At electroweak scale $M_{Z}=91.18GeV$, we have\cite{Zyla:2020zbs} \newline

\tab $\alpha_{S}(M_{z})=\alpha_{3}(M_{z})=0.1179\pm 0.009$ \\

\tab $\alpha_{em}^{-1}(M_{z})=127.952\pm 0.009$ \\

\tab $sin^{2}\theta_{W}(M_{z})=0.23121\pm 0.00017$ \\
The matching condition  at $M_{Z}$ scale gives \\
\begin{equation}\label{e6}
\frac{1}{\alpha_{em}(M_{Z})}= \frac{5}{3}\frac{1}{\alpha_{1}(M_{Z})}+\frac{1}{\alpha_{2}(M_{Z})}
\end{equation}
From the definition of Weinberg mixing angle, we have \\
\begin{equation}\label{e7}
sin^{2}\theta_{W}(M_{Z})=\dfrac{\alpha_{em}(M_{Z})}{\alpha_{2}(M_{Z})}
\end{equation}
where $\alpha_{i}(i=1,2,3)$ are electromagnetic, weak, strong coupling constants respectively and $\theta_{W}$ is the Weinberg angle.\\

Using the values of $\alpha_{3}(M_{Z})$, $\alpha_{em}^{-1}(M_{Z})$,  $sin^{2}\theta_{W}(M_{Z})$ together with eqns.\ref{e6} and \ref{e7} , we have 
\begin{center}
\tab $\alpha_{1,2,3}^{-1}(M_{Z})$=(59.021, 29.584, 8.482) \\
\end{center}
Using the relation $g_{i}=\sqrt{4\pi \alpha_{i}}$  and the values of $\alpha_{i}^{-1}(M_{Z})$ where $i=1,2,3$, the values of the gauge couplings at $M_{Z}$ scale are  
\begin{center}
$g_{1,2,3}(M_{Z})=(0.461469,0.6514018, 1.21740) $ \\
\end{center}
The values of the coupling constants at $m_{t}$ scale can be obtained by using the following one-loop RGE for gauge couplings(Non-SUSY) in the energy scale, $M_{Z}\leq \mu \leq m_{t}$  \cite{bjorkman1985unification} \\
\begin{equation}
\alpha_{i}^{-1}(m_{t})= \alpha_{i}^{-1}(M_{Z})-\dfrac{b_{i}}{2\pi}ln \left(\dfrac{m_{t}}{M_{Z}}\right)
\end{equation}
where $\mu$=$m_{t}$, $i=1,2,3$ and $b_{i}=\left(\dfrac{53}{10}, -\dfrac{1}{2},-4\right)$ for $n_{f}=5$ and $n_{H}=1$\cite{langacker1993uncertainties}\\
With this, we have 
\begin{center}
\tab $\alpha_{1,2,3}^{-1}(m_{t})=(58.482, 29.635, 8.89)$ \\
\end{center}
Correspondingly, 
\begin{center}
\tab $g_{1,2,3}(m_{t})=(0.463547,0.651186,1.189021 )$ \\
\end{center}
The third family charged-lepton masses at $m_{t}$ scale are given by 

$m_{t}(m_{t})=m_{t}(m_{t})$ ;
$m_{b}(m_{t})=\dfrac{m_{b}(m_{b})}{\eta_{b}}$ ;
$m_{\tau}(m_{t})=\dfrac{m_{\tau}(m_{\tau})}{\eta_{\tau}}$ \\

where the physical masses are $m_{t}(m_{t})= 172.76GeV$, $m_{b}(m_{b})=4.18GeV$, $m_{\tau}(m_{\tau})=1.777GeV$ \cite{Zyla:2020zbs} and  $\eta_{b,\tau}$ are QCD-QED rescaling factors with $\eta_{b}\simeq1.53$, $\eta_{\tau}\simeq1.015$\cite{2005Prama..65.1015P}. \\

Now, the values of top quark,bottom quark and tau lepton Yukawa couplings at $m_{t}$ scale with tan $\beta =58$ are obtained by using the relations from MSSM as given below\cite{sashikanta2015effects} \\
\begin{equation}
h_{t}(m_{t})=\dfrac{m_{t}(m_{t})}{v sin \beta}= 0.9930211
\end{equation}
\begin{equation}
h_{b}(m_{t})=\dfrac{m_{b}(m_{b})}{v \eta_{b}cos \beta}=0.910811
\end{equation}
\begin{equation}
h_{\tau}(m_{t})=\dfrac{m_{\tau}(m_{\tau})}{v \eta_{\tau}cos \beta}=0.583666
\end{equation}
where $v$=174GeV is the vacuum expectation value in non-SUSY.
We provide the above values of gauge couplings and  Yukawa couplings as inputs. The values of the gauge couplings and Yukawa couplings at different $M_{R}$ scales are computed using the following RGEs\cite{sashikanta2015effects,Parida_2020,jones1984beta} for gauge couplings and Yukawa couplings(SUSY) in energy scale, $m_{t}\leq \mu \leq M_{R}$ \\
\begin{equation}
\dfrac{dg_{i}}{dt}=\dfrac{b_{i}}{•16\pi^{2}}g_{i}^{3}+ \dfrac{1}{(16\pi^{2})^{2}}\left[\sum_{j=1}^{3}b_{ij}g_{i}^{3}g_{j}^{2}-\sum_{j=t,b,\tau}a_{ij}g_{i}^{3}h_{j}^{2}\right]
\end{equation}
where t=ln $(\mu$/1GeV) ,  $b_{i}$=(6.6,1.6,-3.0);\\
\begin{equation}
b_{ij}=
\begin{pmatrix}
7.96 & 5.40 & 17.60 \\
1.80 & 25.00 & 24.00 \\
2.20 & 9.00 & 14.00 
\end{pmatrix} 
\newline
; a_{ij}=
\begin{pmatrix}
5.2 & 2.8 & 3.6 \\
6.0 & 6.0 & 2.0 \\
4.0 & 4.0 & 0.0
\end{pmatrix}
\end{equation}
\begin{center}
\begin{equation}
\dfrac{dh_{t}}{dt}=\dfrac{h_{t}}{16\pi^{2}}\left(6h_{t}^{2}+h_{b}^{2}-\sum_{i=1}^{3}c_{i}g_{i}^{2}\right)
\end{equation}
\end{center}

\begin{equation}
\dfrac{dh_{b}}{dt}=\dfrac{h_{b}}{16\pi^{2}}\left(6h_{b}^{2}+h_{\tau}^{2}+h_{t}^{2}-\sum_{i=1}^{3}c_{i}^{'}g_{i}^{2}\right)
\end{equation}

\begin{equation}
\dfrac{dh_{\tau}}{dt}=\dfrac{h_{\tau}}{16\pi^{2}}\left(4h_{\tau}^{2}+6h_{b}^{2}-\sum_{i=1}^{3}c_{i}^{''}g_{i}^{2}\right)
\end{equation}
where 
$c_{i}=
\left(\dfrac{13}{15},3,\dfrac{16}{3}\right)$ ; $c_{i}^{'}=\left(\dfrac{7}{15},3,\dfrac{16}{3}\right)$ ;  $c_{i}^{''}=\left(\dfrac{9}{5},3,0\right)$ 

\begin{table}[ht]\label{t1}
\begin{center}
\begin{tabular}{|c| c| c| c|}
\hline
$t_{o}=5.1519$ & $t_{R}=29.93$ & $t_{R}=32.236$ & $t_{R}=34.54$ \\
\hline
$g_{1}=0.463547$ & $g_{1}=0.626687$  & $g_{1}=0.652034$ & $g_{1}=0.680728$ \\
\hline
$g_{2}=0.651186$ & $g_{2}=0.708076$ & $g_{2}=0.713981$ & $g_{2}$=0.720034 \\
\hline
$g_{3}=1.189021$ & $g_{3}=0.783018$ & $g_{3}=0.763223$ & $g_{3}=0.744857$ \\
\hline
$h_{t}=0.993021$& $h_{t}=0.706396$ & $h_{t}=0.680164$ & $h_{t}=0.654667$ \\
\hline
$h_{b}=0.910811$& $h_{b}=0.765436$ & $h_{b}=0.749463$ & $h_{b}=0.733492$ \\
\hline
$h_{\tau}=0.583666$& $h_{\tau}=0.818337$ & $h_{\tau}=0.833217$ & $h_{\tau}=0.848218$ \\
\hline
\end{tabular}
\end{center}
\caption{\label{t1}\footnotesize{Values of gauge couplings and Yukawa couplings at $t_{o}$=ln $m_{t}$, $t_{R}$=ln $M_{R}$ where $m_{t}$=172.76GeV,  $M_{R}=10^{13}$, $10^{14}$,  $10^{15}GeV$.}}
\end{table}

Figs.\ref{f1}- \ref{f2}  show the three gauge couplings and third-generation Yukawa couplings unification at $M_{U}=2\times10^{16}GeV$ and $ 1.25 \times 10^{11}GeV$ respectively. With the values above, we put arbitrary mass eigenvalues($m_{1}$,$-m_{2}$,$m_{3}$) in the range 0.03-0.059GeV(0.04-0.03GeV) for normal(inverted)mass ordering along with the three fixed mixing angles of TBM mixing matrix in one case and alongwith the three fixed mixing angles of GR mixing matrix in the other case as input. Using these several values at different high scales, we obtain the neutrino parameters at low energy by simultaneously solving the RGEs \ref{e1}, \ref{e3}, \ref{e4}, \ref{e5} mentioned above.\newline

At first, we consider neutrino mass eigenvalues following pure normal and inverted mass ordering as input. Here, we find the output values of         sin$\theta_{13}$ beyond the experimentally allowed range even though we vary the $\tan \beta$ value from very small to large value. We consider degenerate NH(IH) in the present work. \newline

In Table  \ref{t2}-\ref{t3} , $m_{i}^{0}$ and $s_{ij}^{0}$ are input values with fixed mixing angles of TBM mixing matrix, at different high scales while $m_{i}$ and $s_{ij}$ ($i,j$=1,2,3; $i\neq j$) are output values at $m_{t}$ scale. For different high scales, we take arbitrary neutrino mass eigenvalues as input but with same mixing angles. If we take same mass eigenvalues for different high scales, then we cannot obtain the experimentally allowed neutrino parameters. In normal hierarchy, we find that $\theta_{12}=35.8^{o}$, $\theta_{13}\approx 8.5^{o}$, $\theta_{23}\approx 51.5^{o}$ while in inverted hierarchy, we find that $\theta_{12}=35.7^{o}$, $\theta_{13}\approx 8.2^{o}$, $\theta_{23}\approx 38.8^{o}$ which is slightly less than that from experimental data. The mass squared differences $\Delta m_{21}^{2}$,$\Delta m_{31}^{2}$,$\Delta m_{32}^{2}$ lie in the allowed range as given by\cite{Esteban_2020} except for the case of $t_{R}=29.93$ (TBM-IH) which is slightly large. Here, we obtain an important result that in NH, $\theta_{23}>45^{o}$ and in IH, $\theta_{23}<45^{o}$. Figs. \ref{f3}-\ref{f4}  represent evolution of mass eigenvalues and mixing angles of TBM mixing matrix with high energy scale in both NH and IH at $t_{R}=29.93$ . For $t_{R}=32.236, 34.54$, the evolutions are almost the same with that of $t_{R}=29.93 $.\\

In Table    \ref{t4}-\ref{t5}, we put arbitrary mass eigenvalues with fixed mixing angles of GR mixing matrix as input for both the normal and inverted mass orderings. Then, we compute the output for different high scales$M_{R}$. In this case also, we find that the neutrino oscillation parameters $\theta_{12}$, $\theta_{13}$, $\Delta m_{21}^{2}$,$\Delta m_{31}^{2}$,$\Delta m_{32}^{2}$ all lie in the experimentally allowed range while $\theta_{23}$ for inverted hierarchy is slightly smaller than that in the experimental data. In NH, we find that $\theta_{23}>45^{o}$ and in IH, $\theta_{23}<45^{o}$ which is similar to what we have found in case of TBM. Figs. \ref{f5}-\ref{f6} represent evolution of mass eigenvalues and mixing angles of GR mixing matrix with high energy scale in both NH and IH at $t_{R}=29.93$ . For $t_{R}=32.236, 34.54$, the evolutions are almost same with that of $t_{R}=29.93$. It is generally observed that both the $\theta_{12}$ and $\theta_{13}$ always magnify with decrease of energy scale unlike $\theta_{23}$. While neutrino masses $m_{i}$ also increase with decrease of energy scale.


\begin{table}[H]\label{t2}
\begin{center}
\begin{minipage}{.85 \textwidth}
\begin{tabular}{|c |c| c| c|c|}
\hline
 TBM - NH & $t_{R}=29.93$ & $t_{R}=32.236$ & $t_{R}=34.54$ \\
\hline
$m_{1}^{0}(eV)$ & 0.04349      &  0.03899    & 0.03399 \\
\hline
$m_{2}^{0}(eV)$ & -0.04479     & -0.04029    & -0.03529\\
\hline
$m_{3}^{0}(eV)$& 0.05873    &  0.05453  & 0.04977 \\
\hline
$s_{12}^{0}$ & 0.5773503   &  0.5773503    &0.5773503 \\
\hline
$s_{13}^{0}$ & 0.0    & 0.0   &0.0  \\
\hline
$s_{23}^{0}$ & 0.707107    & 0.707107   & 0.707107\\
\hline
\hline
$m_{1}(eV)$ & 0.060482   & 0.056103 & 0.050666 \\
\hline
$m_{2}(eV)$ & -0.061126 & -0.056760 & -0.051362\\
\hline
$m_{3}(eV)$  & 0.078716 & 0.075298  & 0.070839 \\
\hline
$s_{12}$ & 0.58443 & 0.58456 & 0.584327 \\
\hline
$s_{13}$& -0.14985 & -0.14981 & -0.145484 \\
\hline
$s_{23}$ & 0.782026 & 0.782830 & 0.781903 \\
\hline
$\Delta m_{21}^{2}(10^{-5}eV^{2})$ & 7.8 & 7.42 & 7.1 \\
\hline
$\Delta m_{31}^{2}(10^{-3}eV^{2})$ &2.538 & 2.52 & 2.451 \\
\hline
$\sum_{i} |m_{i}|(i=1,2,3)$(eV) &  0.20 &  0.188 & 0.173 \\
\hline
\end{tabular}
\end{minipage}
\end{center}
\caption{\label{t2}\footnotesize{Input values ($m_{i}^{0},s_{ij}^{0}$) with fixed mixing angles of TBM mixing matrix   and output values ($m_{i}$,$s_{ij}$) with three different high energy scales, $t_{R}$(=$\ln M_{R}/1GeV$) for normal hierarchy(NH).}}

\begin{center}
\begin{minipage}{.85 \textwidth}
\begin{tabular}{|c|c|c|c|}
\hline
TBM-IH  & $t_{R}=29.93$ & $t_{R}=32.236$ & $t_{R}=34.54$ \\
\hline
$m_{1}^{0}(eV)$ & 0.05232     &  0.04856    & 0.04529 \\
\hline
$m_{2}^{0}(eV)$ & -0.05463     &-0.05095   & -0.04774\\
\hline
$m_{3}^{0}(eV)$&0.04278   &  0.03881  & 0.03535  \\
\hline
$s_{12}^{0}$ & 0.5773503   &  0.5773503    &0.5773503 \\
\hline
$s_{13}^{0}$ & 0.0    & 0.0   &0.0  \\
\hline
$s_{23}^{0}$ & 0.707107    & 0.707107   & 0.707107\\
\hline
\hline
$m_{1}(eV)$ & 0.074070   & 0.07129 & 0.069014 \\
\hline
$m_{2}(eV)$ & -0.074599 & -0.071837& -0.069561\\
\hline
$m_{3}(eV)$  & 0.0562921 & 0.052484 & 0.049161 \\
\hline
$s_{12}$ &  0.5839937  & 0.584154& 0.584323 \\
\hline
$s_{13}$& 0.1433709 & 0.143462 & 0.143469\\
\hline
$s_{23}$ & 0.626464 & 0.625306 & 0.624038 \\
\hline
$\Delta m_{21}^{2}(10^{-5}eV^{2})$ & 7.86 & 7.87 & 7.59 \\
\hline
$\Delta m_{32}^{2}(10^{-3}eV^{2})$ &-2.396 & -2.41 & -2.422 \\
\hline
$\sum_{i} |m_{i}|(i=1,2,3)$(eV) &  0.205 &  0.196 & 0.188 \\
\hline
\end{tabular}
\end{minipage}
\end{center}
\caption{\label{t3}\footnotesize{Input values ($m_{i}^{0},s_{ij}^{0}$) with fixed mixing angles of TBM mixing matrix   and output values ($m_{i}$,$s_{ij}$) with three different high energy scales, $t_{R}$(=$\ln M_{R}/1GeV$) for inverted hierarchy(IH).}}
\end{table}

\begin{table}[H]\label{t4}
\begin{center}
\begin{minipage}{.85 \textwidth}
\begin{tabular}{|c|c|c|c|}
\hline
 GR-NH & $t_{R}=29.93$ & $t_{R}=32.236$ & $t_{R}=34.54$ \\
\hline
$m_{1}^{0}(eV)$ & 0.04354      &0.04058      & 0.03551  \\
\hline
$m_{2}^{0}(eV)$ & -0.04531     & -0.04239  & -0.03735 \\
\hline
$m_{3}^{0}(eV)$& 0.05899  & 0.05633  & 0.05179   \\
\hline
$s_{12}^{0}$ & 0.525739   &0.525739     & 0.525739 \\
\hline
$s_{13}^{0}$ & 0.0    & 0.0   &0.0  \\
\hline
$s_{23}^{0}$ & 0.707107    & 0.707107   & 0.707107\\
\hline
\hline
$m_{1}(eV)$ & 0.060928    & 0.058778  & 0.053327 \\
\hline
$m_{2}(eV)$ & -0.061542 & -0.059402  & -0.054039 \\
\hline
$m_{3}(eV)$  & 0.078950  & 0.077683  & 0.073600 \\
\hline
$s_{12}$ & 0.531808 & 0.532405  & 0.532059 \\
\hline
$s_{13}$& -0.14419 & -0.14997  & -0.14351 \\
\hline
$s_{23}$ & 0.771175 & 0.774186  & 0.77264 \\
\hline
$\Delta m_{21}^{2}(10^{-5}eV^{2})$ & 7.52 & 7.381 & 7.65 \\
\hline
$\Delta m_{31}^{2}(10^{-3}eV^{2})$ &2.52 & 2.57 & 2.573 \\
\hline
$\sum_{i} |m_{i}|(i=1,2,3)$(eV) &  0.201 &  0.196 & 0.181 \\
\hline
\end{tabular}
\end{minipage}
\end{center}
\caption{\label{t4}Input values ($m_{i}^{0},s_{ij}^{0}$) with fixed mixing angles of Golden Ratio(GR) mixing matrix   and output values($m_{i}$,$s_{ij}$) with three different high energy scales, $t_{R}$(=$\ln M_{R}/1GeV$) for normal hierarchy(NH).}

\begin{center}
\begin{minipage}{.85 \textwidth}
\begin{tabular}{|c|c|c|c|}
\hline
GR-IH  & $t_{R}=29.93$ & $t_{R}=32.236$ & $t_{R}=34.54$ \\
\hline
$m_{1}^{0}(eV)$ & 0.05352      &0.05149      & 0.048353  \\
\hline
$m_{2}^{0}(eV)$ & -0.05633     & -0.05445  & -0.05146 \\
\hline
$m_{3}^{0}(eV)$& 0.04449  & 0.04211  & 0.03933  \\
\hline
$s_{12}^{0}$ & 0.525739   &0.525739     & 0.525739 \\
\hline
$s_{13}^{0}$ & 0.0    & 0.0   &0.0  \\
\hline
$s_{23}^{0}$ & 0.707107    & 0.707107   & 0.707107\\
\hline
\hline
$m_{1}(eV)$ & 0.076043   & 0.075905  & 0.074059 \\
\hline
$m_{2}(eV)$ & -0.076558  & -0.076367& -0.0745402 \\
\hline
$m_{3}(eV)$  & 0.058607 & 0.057011  & 0.054739 \\
\hline
$s_{12}$ & 0.53185 & 0.532192 & 0.53299 \\
\hline
$s_{13}$& 0.14334 & 0.145793  & 0.15349  \\
\hline
$s_{23}$ & 0.63675 & 0.63459 & 0.62987\\
\hline
$\Delta m_{21}^{2}(10^{-5}eV^{2})$ & 7.86 & 7.04 & 7.14 \\
\hline
$\Delta m_{32}^{2}(10^{-3}eV^{2})$ &-2.43 & -2.57 & -2.56\\
\hline
$\sum_{i} |m_{i}|(i=1,2,3)$(eV) &  0.21 &  0.209 & 0.203 \\
\hline
\end{tabular}
\end{minipage}
\end{center}
\caption{\label{t5}Input values ($m_{i}^{0},s_{ij}^{0}$) with fixed mixing angles of Golden Ratio(GR) mixing matrix   and output values ($m_{i}$,$s_{ij}$) with three different energy scales, $t_{R}$(=$\ln M_{R}/1GeV$) for inverted hierarchy(IH).}
\end{table}

\newpage
From the tables above, it is seen that both the case of NH and IH of TBM and GR mixing matrix are valid at high energy scale considering the bound $\sum_{i}|m_{i}|<0.23$eV. This result contradicts the claim in earlier analysis \cite{zhang2016viability}, although the analysis is discussed with the inclusion of CP violating phases.

\section{Conclusion and Discussion}
We have presented a detailed analysis of radiative corrections of neutrino masses and mixings in MSSM with TBM mixing and GR mixing at three values of high energy $M_{R}$ scales $10^{13}$,$10^{14}$,$10^{15}$ GeV. In all cases, we have used large value of tan $\beta$=58. We take arbitrary values on the three neutrino mass eigenvalues at high energy scales for both NH and IH, and obtain the experimentally allowed neutrino oscillation parameters at low energy scale, $m_{t}$. The neutrino mass eigenvalues $|m_{i}|($i=1,2,3$)$ in the range (0.03-0.079) eV provide correct values of the neutrino oscillation parameters at low energy scale. We have successfully obtained almost all the mixing angles and mass squared differences in the 3$\sigma$ range of NuFit data. The value of the sum of absolute neutrino masses, $\sum_{i}|m_{i}|$ is in the range (0.17-0.20) eV for normal hierarchy and (0.188-0.21) eV for inverted hierarchy. These values are within the upper bound on $\sum_{i}|m_{i}|$$<$ 0.23eV at the 95\% confidence level from Planck 2015\cite{Planck:2015fie}. However, these values are still higher as compared with  the latest Planck bound $\sum_{i}|m_{i}|$$<$ 0.12eV\cite{Planck:2018vyg,PhysRevD.103.083533,de_Salas_2021}. There are a lot of uncertainties in the process of observation of cosmological bound on the sum of the absolute three neutrino masses\cite{https://doi.org/10.48550/arxiv.2112.02993,PhysRevLett.123.081301}. It may still requires further analysis to achieve the most correct cosmological bound in future measurements. \newline

\tab In the present analysis, we have not included the effects of running of the Dirac and Majorana CP phases. We also consider $m_{t}$=172.76GeV as the SUSY breaking scale $m_{s}$ for simplicity. The analysis shows the validity of the TBM and GR mixings at high seesaw scale $M_{R}$, and the deviations are the effects of radiative corrections at low energy scale.  For future investigations, it would be very interesting to consider higher SUSY breaking scale,$m_{s}$  as well as higher $M_{R}$ scale($2\times 10^{16}$GeV) for different $tan \beta$ values \cite{PhysRevD.97.055038}.
\newpage 
{\begin{figure}[H]

\includegraphics[scale=0.43]{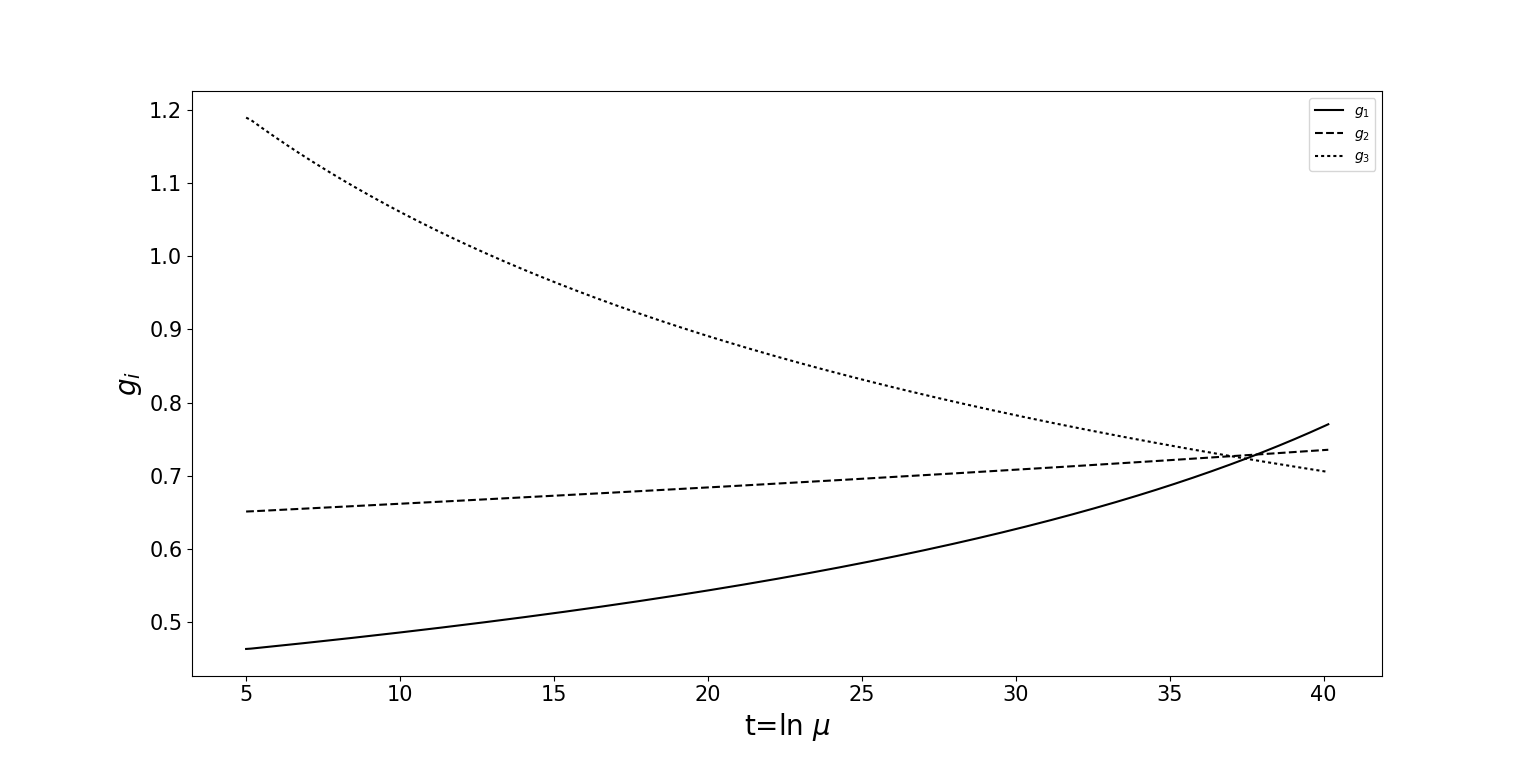}\caption{\footnotesize Three gauge couplings unification is observed at $M_{U}=2\times 10^{16}$GeV.}\label{f1}

\includegraphics[scale=0.43]{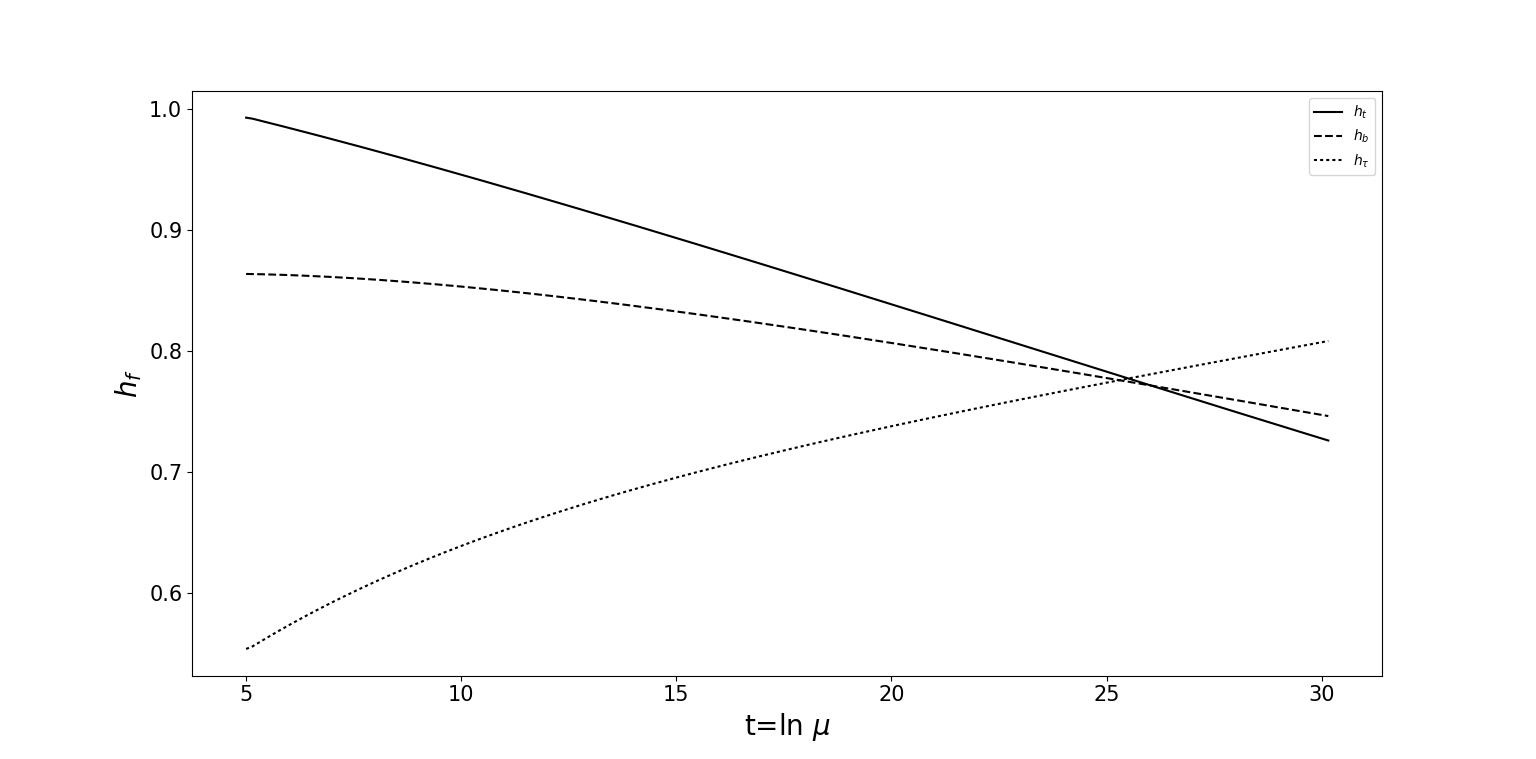}\caption{\footnotesize Third-generation Yukawa couplings unification is observed at $\mu=1.25\times10^{11}$GeV.}\label{f2}

\end{figure}
}

\newpage
{
\begin{figure}[htbp]

\centering
\includegraphics[width=165mm]{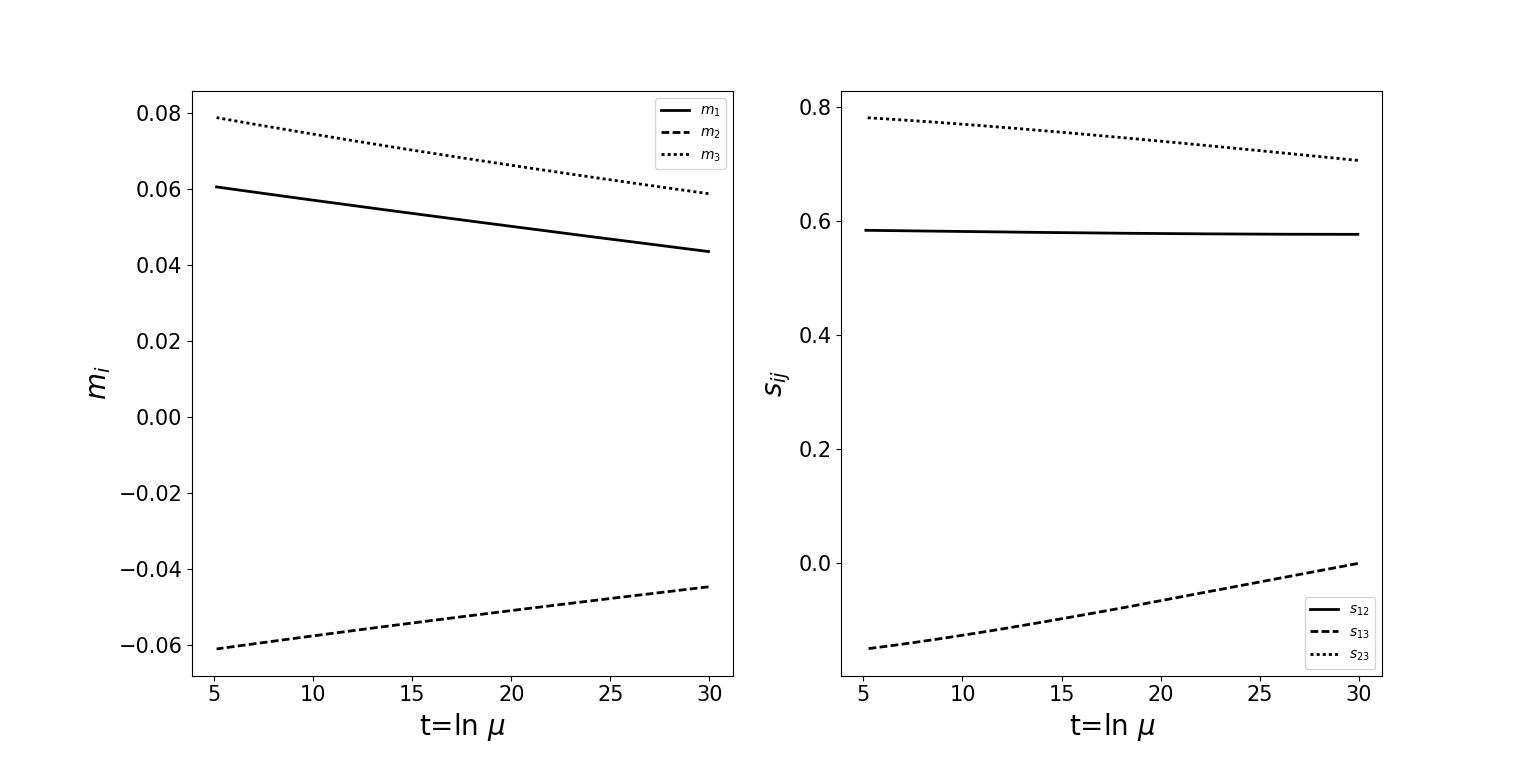}\caption{\footnotesize Evolution of mass eigenvalues and mixing angles of TBM mixing matrix with energy scale in NH for $t_{R}=29.93$.}\label{f3}
\includegraphics[width=165mm]{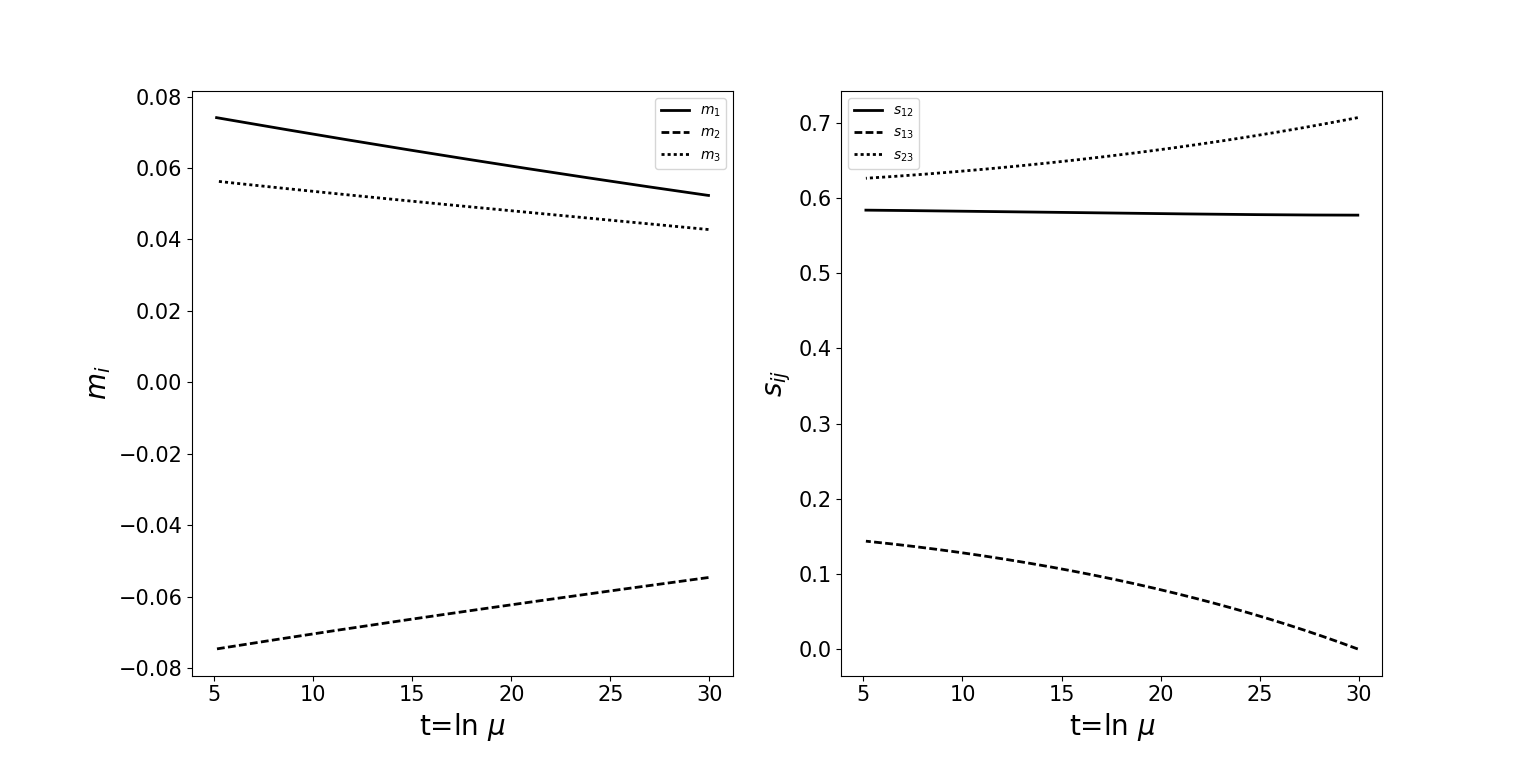}\caption{\footnotesize Evolution of mass eigenvalues and mixing angles of TBM mixing matrix with energy scale in IH for $t_{R}=29.93$.}\label{f4}
\end{figure}
}
\newpage

{\begin{figure}[htbp]
\includegraphics[width=165mm]{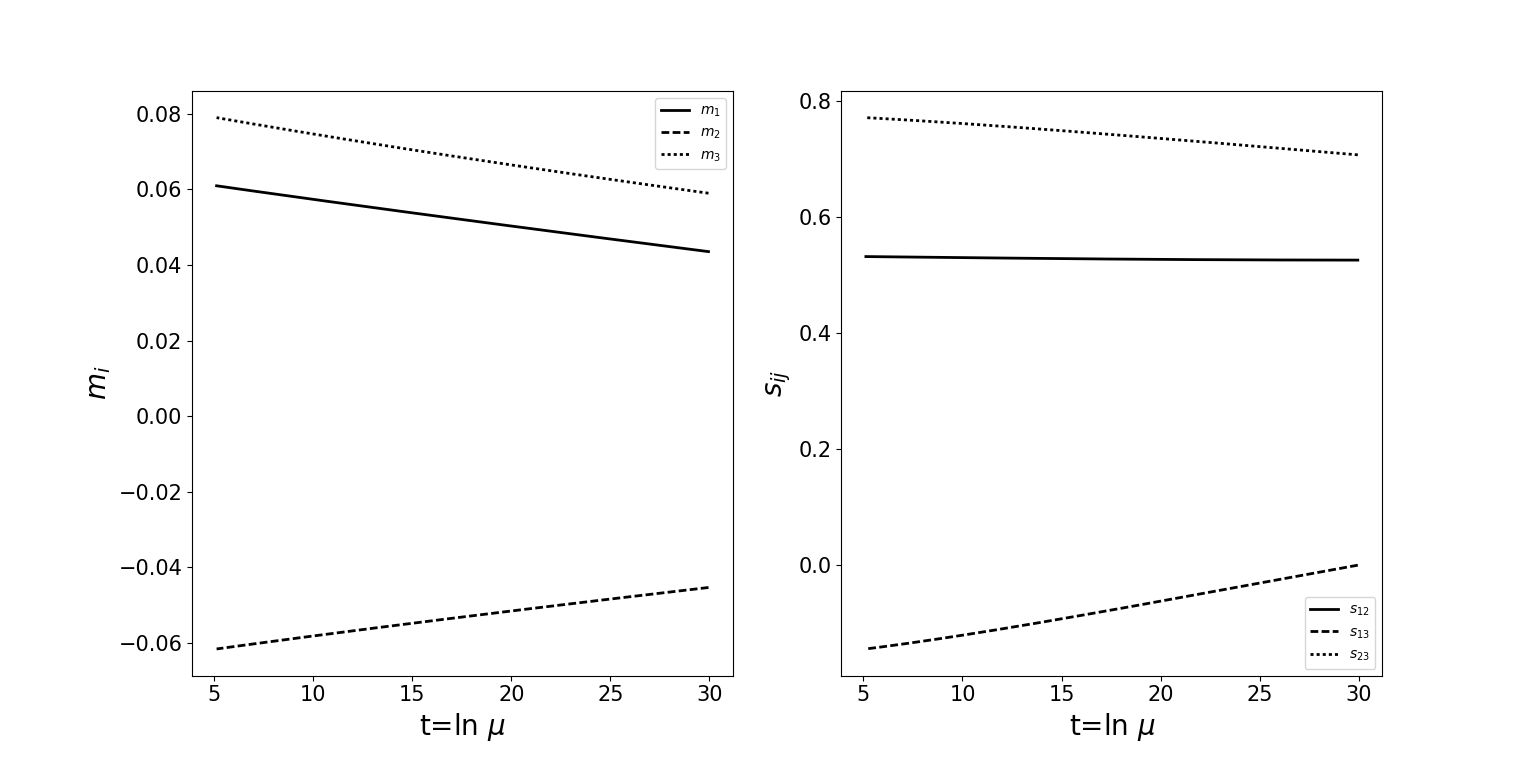}\caption{\footnotesize Evolution of mass eigenvalues and mixing angles of GR mixing matrix with energy scale in NH for $t_{R}=29.93$.}\label{f5}

\includegraphics[width=165mm]{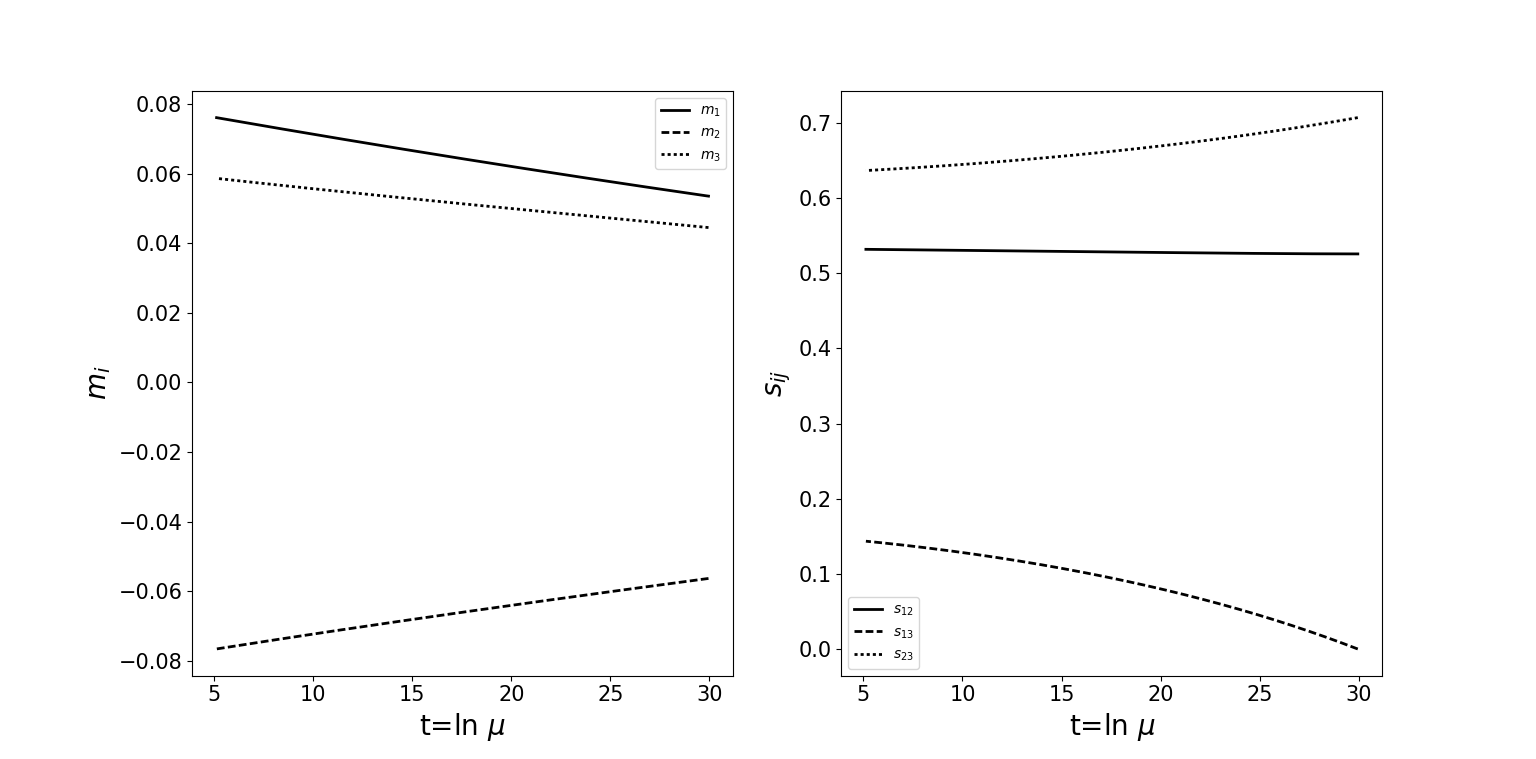}\caption{\footnotesize Evolution of mass eigenvalues and mixing angles of GR mixing matrix with energy scale in IH for $t_{R}=29.93$.}\label{f6}
\end{figure}
}

\clearpage

\section{Acknowledgement}
One of us (PW) would like to thank Manipur University for granting  Fellowship for Ph.D. programme. 
\bibliographystyle{unsrt}
\bibliography{newref}

\end{document}